# IMPACTS!

## A program to calculate the effects of a hypervelocity impact into a Solar System body

Keith A. Holsapple
PO Box 305
Medina, WA. 98039
kholsapple@comcast.net

## Abstract

IMPACTS! is a free web based Javascript application that can be opened and executed in any standard browser. It can be downloaded from https://www.dropbox.com/sh/2oanw5vxf09qpl1/AAD8uD-CsWir6slPlWndyUQAa?dl=0. This program calculates the many characteristics of the craters, ejecta and disruptions that are created from hypervelocity impacts into the objects of the Solar System. The code cam also be obtained from the author.

## 1. Introduction

The calculations are defined by an input window that defines the target and the impactor. Target types include lunar, terrestrial, laboratory, planetary or an asteroid or planet satellite. The target properties for each are suggested but can be changed by the user. The impactor is defined by its size, impact velocity, impact angle, and composition.

The user-chosen defining conditions determine the crater morphology. Those may be the smallest, spall craters in brittle targets; the simple bowl-shaped strength-controlled craters; the bowl-shaped gravity- controlled craters; and the large lunar and terrestrial complex craters including flat floors, peak rings and multiple rings. The calculated results include crater dimensions, melt and vapor amounts, ejecta characteristics, formation time and the standard non-dimensional 'Pi' groups. Further, for asteroids, the catastrophic dispersion conditions are given. All variables, both inputs and outputs, can be specified in any choice of units, the program makes all unit conversions automatically. In addition to the individual output values, six types of plots can be displayed.

The results are based on the physically based scaling methods developed over the last 30 years, by Keith Holsapple, Robert Schmidt and Kevin Housen, as indicated in the reference list below. It does not use older and now discredited methods such as "Energy Scaling" or strictly semi-empirical dimensional forms. Instead, it is based on the non-dimensional forms required by any valid physical theory. But it also uses an important simplification, the point-source approximations, that have been validated by 30 or more years of application, by experiments, field data, and numerical simulations.

The nature of that point source approximation can be described. It is applicable for physical processes in given materials in which the 'source' occurs in a very small region and over a very short time, compared to the final problem dimensions and duration. That is indeed the case for a hypervelocity impact. In that case, the results at any distance and time scale much larger than the source will not depend upon the details of that source. They cannot depend upon any source measures with a timescale nor a length scale. Instead, they can only depend upon one single power law combination of the primary source measures. In continuum mechanics there are only three such measures corresponding to the three basic balance equations of mass, momentum, and energy. One such choice is the power law combination $aU^\mu \delta^\nu$ of the source dimension $a$, its velocity $U$ and its mass density $\delta$, with scaling powers of $\mu$ $and$ $\nu$. That combination was first called a 'coupling parameter' in Holsapple and Schmidt, JGR 1987. Equally, any other single measure that can be obtained by these three such as powers of mass, mass density, and energy and can be used.

Using that assumption, together with dimensional analysis, the scaling theory supplies the necessary functional forms for all of the dependences on the problem inputs, but not the necessary scaling exponents nor the leading coefficients. For those, one needs experiments. And there are many lab impact tests, and some numerical code calculations (although we must be wary, some of the codes do a poor job of describing the material properties). In addition, there are known similarities (Holsapple, 1980) between explosive and impact cratering, so the explosive database gives importance additional guidance to the laboratory tests and numerical simulations of impacts.

While these scaling theories are quite highly developed, they are not always easily accessible to a researcher. Results are often couched in "dimensionless groups", with so-called "pi-groups" that are sometimes difficult to unravel. Further, as is true for any study in science or engineering, it is easy to get the units wrong. In addition, the actual data that provides the quantitative backbone to the scaling theories is scattered throughout the literature; and much of the data for explosives is in the so-called "grey literature" of military and company reports that is not always easily accessible to a general researcher.

The purpose of this paper is to make that data and theory easily accessible to any person with a scientific background; using a web-based application that allows the user to define conditions of interest and obtain all the primary characteristics of a resulting crater for impacts.

## 2. Theory

For impacts the "source" of an event is the impactor. We assume that the source is much smaller than a "target" body. The impact velocity is assumed to be greater than the target sound speed: from a few to many km/s. The source shape is of little consequence if it is relatively compact. The results do not apply to long rod or penetrator impactors, nor to slower m/s impact velocities. The results do account for different impact angles, but not extremely shallow, glancing impacts.

The cratering outcomes for a given source are determined by the source and by the target strength and the surface gravity, and often only one of theselatter two. The smaller craters are determined by the target strength, while the larger craters are determined by the surface gravity.

That defines two regimes of cratering: the "strength regime" and the "gravity regime"[1], with in-between transition cases. Recently, a further subset of the strength regime for the smaller craters has been identified as the spall crater regime (Holsapple, 2022, in preparation).

## 2.1. Dimensionless Forms

As the primary example, the volume $V$ of a crater formed by a given impact can be expected to depend on the impactor radius $a$, velocity $U$, and mass density $\delta$. Note that those 3 variables also defined the kinetic energy, momentum, and mass of the impactor, so equally those could be used, as well as any other choice of three independent variables containing the three independent units of mass, time and length.

The target has some strength measure $Y$, a mass density $\rho$, and a porosity $\phi$. Below it will be assumed that the strength depends on the event size, but for now it is treated as a constant. We can ignore the additional properties for a fixed choice of target types, and we consider only one target type at a time. Then the crater volume depends on those variables by some functional relationship:

$$V = F[\{\rho, g, Y\}, \{a, U, \delta\}] \tag{1}$$

grouped according to target mass density $\rho$, gravity $g$, and (some) strength $Y$ variables and impactor radius $a$, velocity $U$, and mass density $\delta$ variables.

In addition, it will next be assumed that the impactor measure can be considered as a "point source", since the region of influence is much larger than the impactor dimensions. Then we assume that

$$V = F[\{\rho, g, Y\}, aU^\mu \delta^\nu] \tag{2}$$

Finally, as is the case for any physical results, the results can always be stated in terms of dimensionless forms. That is a simply the condition that the results must be independent of the choice of physical units. So, we make a simple dimensional analysis. Definite power-law forms are the outcome of that assumption. The presentation here follows the Holsapple, 1993 review paper on impact cratering as well as Holsapple and Schmidt, 1980, 1982 and 1987. I will not reproduce those presentations here, but will just present the final forms.

The general form for a scaled volume is given in the Eq. 18 of (Holsapple, 1993) is

$$\pi_V = K_1 \left\{ \pi_2 \left(\frac{\rho}{\delta}\right)^{\frac{6\nu-2-\mu}{3\mu}} + \left[K_2 \pi_3 \left(\frac{\rho}{\delta}\right)^{\frac{6\nu-2}{3\mu}}\right]^{\frac{2+\mu}{2}} \right\}^{\frac{-3\mu}{2+\mu}}$$

$$\text{with } \pi_V = \frac{\rho V}{m}, \quad \pi_2 = \frac{ga}{U^2}, \quad \pi_3 = \frac{Y}{\rho U^2} \tag{3}$$

having the two scaling exponents $\mu$ and $\nu$, and two coefficients $K_1$ and $K_2$. Those must be determined from actual data. (The last exponent $-3\mu/(2+\mu)$ was often denoted as $\alpha$.) The first term with $\pi_2$ dominates for large events, and that defines the gravity regime. That dimensionless $\pi_2$ group is sometimes called the "gravity-scaled size parameter". The second $\pi_3$ term with the target strength dominates for small events, it defines the strength regime.

---

[1] For very small target bodies such as asteroids, one should be very careful in assuming the strength level, because even almost negligible cohesion values can dominate the equally negligible gravity.

The $K_1$ constant determines the size of the crater in both the gravity and strength regimes, it is determined by the "early-time" coupling of the source energy and momentum into the target. The $K_2$ constant determine the event size for the transition between strength scaling and gravity scaling.

These exponents and coefficients depend upon the target properties, but here I make no attempt here to characterize those dependences. Also, this equation has no specific inclusion of the angle of repose of the target. The exponents and constants $K_1$ and $K_2$ are individually chosen for each material type from the data.

It is well established over the last several decades (e.g. Holsapple, 1993) that for relatively dissipative materials such as "dry" soils and sands the exponent $\mu$ is about 0.4, and for wet and rocky targets it is about 0.55. (A source couples more kinetic energy into the less porous materials.) The exponent $\nu$ is 1/3 if some combination of the mass and specific energy (defining also the energy and momentum) of the source determines its measure, but experiments give uncertain values, ranging from about 0.2 to 0.4. Here the value of 1/3 is adopted, primarily because of its simplicity. That value is not of much consequence for the ranges of the density ratio $\rho/\delta$ of interest.

And there is one final complexity. It is now generally accepted that the strength of rocky bodies (but not soils and sands) depends on the size scale of the event (e.g. Housen and Holsapple, 1999). The explosion data in the strength regime for rocks clearly shows that affect. The strength one measures on the lab for a 10 cm pristine sample of say, basalt, is not the strength that governs a 100 m crater in basalt. That size dependence of strength is a consequence of the fact that natural geological materials are permeated by distributions of cracks and flaws of all sizes, and it is those flaws that limit the strength.

This web application includes a size-dependent strength, assuming the strength decreases as the negative $1/n$ power of the crater diameter. Specifically, it uses the input of a lab-sized cohesive strength $Y_0$, and then it iterates the strength $Y$ according to the size of the resulting crater using the formula $Y=Y_0(10cm/D_{cm})^{-1/n}$. Thus, the reference "lab" strength $Y_0$ is what is measured for a 10 cm specimen. The degraded value found is indicated by the application.

Those flaws have a much greater effect on tensile failures than on shearing failures. While that feature is not yet well researched, here the size-dependence on the crater sizes, which for normal craters are dominated by the shear strength, use $n=4$. Spall craters are formed primarily by tensile failures, and in that case this program uses $n=2$, which is the smallest expected value for $n$ and which occurs for "fully cracked" materials. The interest reader can consult Housen and Holsapple, 1999.

# 3. Impact Results

## 3.1. Target Materials

In Holsapple 1993, results are given for impacts into each of five target types: dry sand, dry soils, wet soils, dry soft rocks, and hard rocks. That choice of material classifications mimics that for the extensive study of small and large explosions in the Schmidt et al, 1988 DNA report. Those explosive results defined the strength asymptote for small craters and, in most cases, the gravity regime for large ones. To those classifications I added three more: lunar regolith, cold ice, and water which may be of interest in limited cases.

While there are also considerable experimental data for impacts into various material types (mostly lab data), there are also significant gaps in that data. Here I summarize the nature of impact data. What are needed are values for the two constants $K_1$ and $K_2$. Any gravity regime data can be used to obtain $K_1$. Then according to the argument above, using $K_2=1$, the value for the "cratering strength" $Y$ is determined by the strength regime results.

### 3.1.1. Dry Sand

Dry sand targets are often used for laboratory experiments. Both 1G and gravity at up to 500G tests[2] have been made, for velocities from 1 to 5 km/s. Since there is no material strength measure $Y$ for dry sand[3], all results are in the gravity regime. The results conform to the power-law expectations almost exactly and furnish irrefutable evidence of the accuracy of the point-source assumption. This data is primarily for Ottawa sand and is very robust, although one must be aware that different sands do have slightly different cratering results. Those results give guidance to other material's results that are not so well known.

### 3.1.2. Wet Sand

The database includes about a dozen experiments in wet sand, at various gravity levels. Those results were presented in Schmidt and Housen, 1987 and interpreted in terms of gravity and strength regimes. Since the data were mostly in the strength regime, those authors made estimates for a gravity regime based on the water and dry sand results. I use those estimates here.

---

[2] The reason for testing at increased gravity is because it is a way to vary the $\pi_2$ parameter, in place of increasing the impactor size. See the references.
[3] Dry sand gets its strength when subjected to pressure via its angle of friction.

### 3.1.3. Wet Soils

There are no impact experiments for general wet soils, and there may well be significant variation depending on the soil. However, there are some explosive results for wet soils. Here I assume generic results as the same as for the wet sand impacts, and check against the comparisons to the explosive results.

### 3.1.4. Dry Soft Rocks, Hard Rocks, Ice

There are no definitive impact experiments for dry soft rocks, hard rocks or for ice. While there are some small-scale impact experiments, the outcomes are shallow surface spall craters, not excavation craters. It is known that craters in brittle materials at Earth's gravity will be spall craters as long as they are under a few meters in diameter (Holsapple and Housen, 2013), so those small-scale results are not definitive for larger excavation craters. Such spall craters are included as possible outcomes in the application. However, there are large explosive excavation data, but still all in the strength regime. For impacts in the strength regime, I will base the estimates on the explosive data, using the equivalences to impacts.

For the gravity regime, the strength is no longer of any consequence. For that reason, our best estimates are that all non-porous materials have the same gravity regime outcomes as wet sand.

### 3.1.5. Water

Several researchers have reported hypervelocity impacts into water targets. For such impacts, "crater size" is measured when the crater is at its maximum depth. Subsequently the crater collapses, with the radius moving in an outward wave, and the crater center rebounding upwards, creating a centered water-spout. As for dry sand, there is no strength regime, only a gravity one. The scaling of these craters was given in Holsapple and Schmidt, 1982. That interpretation was directly used here.

### 3.2. The Impact Target Constants

Based on the considerations above, the constants used for impacts in this web application are:

| Material | $K_1$ | $K_2$ | $\mu$ | $\nu$ | Y (dynes/cm$^3$) At lab scale | $\rho$ (g/cm$^3$) |
|---|---|---|---|---|---|---|
| Water | 1.0 | 0 | 0.55 | 0.33 | 0 | 1 |
| Dry Sand | 0.15 | 0 | 0.4 | 0.33 | 0 | 1.7 |
| Dry Soil | 0.15 | 1. | 0.4 | 0.33 | 3.e6 | 1.7 |
| Wet Soil | 0.065 | 1. | 0.55 | 0.33 | 3.5e6 | 2.1 |
| Soft Dry Rock/Hard Soils | 0.04 | 1. | 0.55 | 0.33 | 1.3e7 | 2.3 |
| Hard Rocks | 0.06 | 1. | 0.55 | 0.33 | 1.44e8 | 3.0 |
| Lunar Regolith | 0.15 | 1. | 0.4 | 0.33 | 1.e6 | 1.5 |

| Cold Ice |  | 1.93 | 1. | 0.55 | 0.33 | 1.5e8 |  | 0.93 |

Table 1.  Scaling Constant for Impacts

### 3.3. The Impactor and Gravity Properties

For hypervelocity impacts of compact projectiles, the mass density is needed, but not other properties.  Standard values are given for the listed impactor materials.  A user may pick other values in the app are made by choosing "Other" in the pull-down menu for the impactor type.

Gravity is pre-set for Terrestrial, Laboratory and Lunar cases and calculated for the asteroid cases.  Again the input can be adjusted by the user if "Other" is selected.  The velocity can be set to any value, but a warning ensues for any values below 1 km/sec, where the data is sketchy and the point source assumption becomes iffy.  For non-vertical impacts, the vertical component U cos(θ) is used.

### 3.4. Simple Crater Shapes

In each material type, the craters except for the very largest are assumed to have the same fixed "bowl" shape.  The shapes of simple craters are calculated from $R = K_r V^{1/3}$ and the depth from $D = depth = K_d V^{1/3}$.  The values indicated by the data and programmed are as follows:

| **Material** | $K_r$ | $K_d$ |
|---|---|---|
| Water | 0.8 | 0.75 |
| Dry Sand | 1.4 | 0.35 |
| Dry Soils w/ cohesion | 1.1 | 0.6 |
| Soft Rock | 1.1 | 0.6 |
| Hard Rock | 1.1 | 0.6 |
| Cold Ice | 1.1 | 0.6 |

Table 2.  Crater Shape Parameters.

In all cases, the rim diameter is assumed to be 1.3 times the excavation diameter and the lip height 0.36 times the rim diameter, consistent with the data and measured terrestrial and lunar simple craters. The ejecta volume is assumed to be 80% of the excavation volume. The crater formation time is from Schmidt and Housen 1987, and the Figure 12 in (Holsapple, 1993) as

$$T = 0.8\sqrt{\frac{V^{1/3}}{g}} \qquad (4)$$

### 3.5. Melt and Vapor Volumes

Melt and vaporization of target material occurs when the initial impact pressure is high enough. That is defined by the equation of state; it is where the Hugoniot curve crosses the melt boundary.  For melt, I assume that the velocity threshold is $U = \sqrt{10 E_{melt}}$ in terms of the specific energy of melt for the material.   I take a generic value for the melt energy for silicates as 5 $10^{10}$ ergs/g. I use the "less than energy scaling" from Holsapple, 2003 matched to some of the results from (Pierazzo et al., 1997) and get

$$V_{melt} = 0.5 V_{impactor} \left[\frac{U^2}{5E10} - 10\right]^{0.9} \quad (5)$$

Vapor production is in a volume much closer to the impactor, so I use strict energy scaling with a generic vapor energy of 1.5E11:

$$V_{vapor} = 0.4 V_{impactor} \left[\frac{U^2}{1.5E11} - 10\right]^{1.0} \quad (6)$$

I have not yet added the melt and vapor for impacts into ice, there are significant questions about its many phases at cold temperatures.

### 3.6. Complex Craters

For craters with a simple transient radius greater than some value $R^*$, the simple excavation crater with the radius $R_e$ undergoes a late-time readjustment into a much broader and shallower "complex crater". The data for lunar craters by Pike 1977 gives a transition to complex shapes beginning at $D^*=10.6$ km rim diameter. The transition in rim heights begins at a larger size, $D^*_{rh}=22.8$ km diameter. The onset of flat floors is gradual, but is fully developed at $D^*_{fl}=20$ km diameter. On the earth, those transition diameters are reduced by the inverse of the gravity levels (Pike, 1977).

Consistent with general crater scaling, those transition values undoubtedly depend on the ratio of lithostatic pressure at the appropriate size scale to the strength at that same size scale. However, we really don't have definitive information about appropriate strengths. Therefore, whatever strength is chosen in the application, I assume that these transition diameters are at these same values for complex craters.

Let $D_r^f$ denote the final rim diameter, and $D_r^t$ the transient (simple) rim diameter. The analysis of the relation between simple and complex craters is based on an incompressible readjustment from the simple crater shapes measured in laboratory experiments and those observed for lunar craters, using primarily the data of Pike 1977. The approach is outlined in Holsapple, 1993. The primary result is an expression for the ratio of the final to transient rim radius:

$$\frac{D_r^f}{D_r^t} = 1.02 \left(\frac{D_r^f}{D^*}\right)^{0.079} \quad (7)$$

which gives, using the ratio 1.3 for the transient rim to excavation rim diamters,

$$D_r^f = 1.33 (D_e)^{1.086} (D^*)^{-0.086} \quad (8)$$

The Pike data for lunar craters gives for the depth of complex craters larger than $D^*=10.6$ km as $d=1.044(D_r^f)^{0.301}$ in km units. This matches the simple crater result, $d = 0.2D_r^f$ at the transition onset using the dimensionally consistent form

$$d = 0.2D^*\left(\frac{D_r^f}{D^*}\right)^{0.301} \tag{9}$$

For the rim height, Pike gives $h= 0.236\,(D_r^f)^{0.399}$ for complex craters and $h = 0.036D$ for simple craters. With the transition of rime heights at $D^*_{rh} =22.8$ km diameter, that gives the equation

$$h = 0.036D^*\left(\frac{D_r^f}{D^*_{rh}}\right)^{0.399} \tag{10}$$

The flat floor diameter is given for lunar complex craters as $D_f=0.187(D_r^f)^{1.249}$ for diameters greater than $D^*_{fl}=20$ km. Assuming this dimension is zero at the 10.6 km onset of complex craters, the fit used was

$$D_f = 0.292\left(D^*_{fl}\right)^{-0.249}\left(D_r^f - D^*\right)^{1.249} \tag{11}$$

Finally, the volume below the rim uses a profile with a flat floor, and a uniform slope from the floor diameter to the rim diameter and the rim height. The outcome is given as

$$vol = \frac{\pi d}{4}\left[D_f^2 + \frac{1}{3}(D_r^f - D_f)(D_r^f + 2D_f)\right] \tag{12}$$

Note that in the app, the display section for complex craters only appears when the crater sizes are larger than the transition diameter.

### 3.7. Ejecta Scaling

The definitive references on the amounts and properties of the ejecta from impact cratering are Housen et al., 1983 and Housen and Holsapple, 2011.

The ejecta characteristics can be described by a plot of ejecta mass with velocity greater than some velocity v versus that velocity v. Such a curve using the PS scaling theory is shown in Fig. 9 (from Housen and Holsapple, 2011). The ejected mass velocity varies from a largest $v_L$ launched from near the impact point to a slowest $v^*$ launched later from the crater rim. Between those values, based on the PS scaling, the mass with velocity greater than the velocity v is given

$$M(> v) = M^*\left(\frac{v}{v^*}\right)^{(-3\mu)}. \tag{13}$$

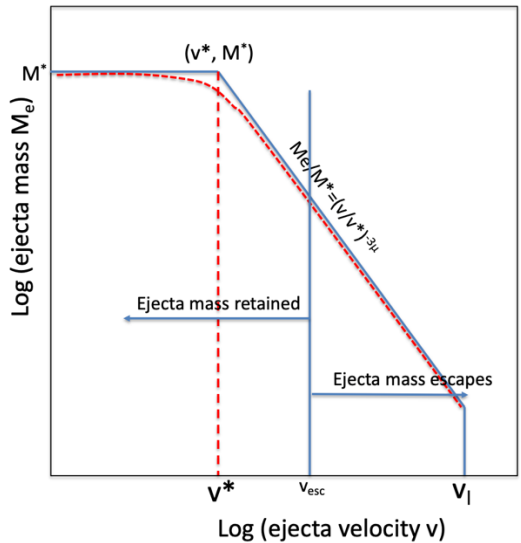

Figure 1. The amount of mass with ejecta velocity greater than some value v. The primary sloped portion is a power-law decreasing with increasing velocity from the coordinates (v*, M*) down to the maximum velocity $v_L$. All of the mass has a velocity greater than v*, thus M* is the total ejecta mass. All ejecta with a velocity greater than the target's escape velocity $v_{esc}$ will escape from the target body; that with smaller velocity will fall back to the target surface and will not contribute to mass loss nor spin-up. The escape velocity depends upon the target size and can be smaller than v*, in which case all ejecta escapes, or larger than $v_L$ in which case no ejecta escapes.

M* is the total ejecta mass which, according to the PS theory, is always some fixed proportion of the crater mass; the remainder of the crater volume is due to compaction, surface uplift and mass movement. That ejecta mass is be estimated as about 60% of the crater mass[4]. The value of the slowest ejecta velocity v* is found from laboratory experimental results for ejecta (e.g. Housen and Holsapple, 2011). For asteroids, it has the values given in Holsapple and Housen 2012, Eqs. (27):

$$v^* = K_{vs}\sqrt{\frac{Y}{\rho}}, \quad \text{or} \quad v^* = K_{vg}U(\pi_2)^{\frac{1}{2+\mu}}, \qquad (14\ a,\ b)$$

in the strength and gravity regimes, respectively. These two additional ejecta scaling constants are then found from experimental data. Here $K_{vs}$ and $K_{vg}$ =0.3 for both asteroid composition types (Holsapple and Housen, 2012). These equations and the value for the escape velocity define the amount of crater ejecta that will escape an asteroid.

It is the smaller impacts governed by the cratering and ejecta theory there are the most numerous. However it is the larger catastrophic and near-catastrophic impacts which create most of the mass loss and erosion.

## 4. Spall Craters

High-speed impact craters formed in rocks, ice and other brittle materials in the lab consist of an outer, broad shallow concentric region formed by tensile fracture (spall), surrounding a smaller central "pit" crater of greater depth. On the Earth, that "spall crater" morphology also occurs for smaller craters formed experimentally by explosions but ceases to exist for craters greater than a few meters in diameter. Further, it is not commonly recognized for craters in the solar system, but it is an issue for cratering on the small brittle asteroids.

---

[4] This directly connects the ejecta mass to the crater mass. The remainder is due to subsurface movement of material: downward in the crater center and upward in the crater rim and beyond. In Holsapple and Housen, 2012 the ejecta mass was found indirectly from measured ejecta results, not directly from crater mass. I now think this present simpler approach is best.

The complete theory will be given in Holsapple, 2022 ( in preparation). The spall craters are formed when

$$\boxed{\text{Spall Crater Regime, Size depend strength} \\ \frac{D}{D_0} < \left[ 0.1 \frac{Y_0}{\rho g D_0} \right]^{\frac{n}{n+1}}} \quad (15)$$

which is always a subset of the strength regime. For spall it is assumed that *n*=2.

## 5. Catastrophic Dispersions

The impactor energy per unit target mass required to either disrupt and or disrupt and disperse the target is denoted Q*. Here the only concern is for that to disrupt and disperse the target so there is no need for further subscripts. The reader is referred to Holsapple and Housen, 2019 for the theory. It is used directly here. However, since that reference considers only the average velocities of 5.5 km/s and impact angles of 45 degrees, corrections must be made for impacts at other angles.

For a given target mass *M* and dimension *R*, and impactor and target mass densities, the scaling theory (e.g. Housen and Holsapple, 1990) shows that impactor mass $m^*$ for a normal impact depends on the impact velocity as $m^* \sim U^{-3\mu}$. Therefore, I assume that for an impact at velocity *U* and angle $\phi$ from the normal direction, the required mass is given as $m^* \sim (Cos\phi\ U)^{-3\mu}$. As a consequence, for a fixed target,

$$Q^* = \frac{m^* U^2}{2M} \sim U^{2-3\mu} (\cos\phi)^{-3\mu}$$

and to go from a velocity 5.5 km/s and angle of 45° to any others, the required specific energy is given as

$$Q^*(\phi, U) = Q^*(45°, 5.5 \tfrac{km}{s}) \left(\frac{U}{5.5}\right)^{2-3\mu} \left(\frac{\cos\phi}{\cos 45°}\right)^{-3\mu} \quad (16)$$

with velocity *U* in units of km/s.

https://www.dropbox.com/sh/2qr68u6ymqotfr2/AADMuT2-sP8d-Lf4oydIdSeSa?dl=0